\documentclass[aps,prl,superscriptaddress,reprint]{revtex4-1}

\usepackage{graphicx}
\usepackage{dcolumn}
\usepackage{bm}
\usepackage{amsmath,amssymb}

\usepackage[utf8]{inputenc}
\usepackage[T1]{fontenc}
\usepackage{mathptmx}
\usepackage{mathrsfs}
\usepackage{bm}
\usepackage{tikz}

\begin{document}

\title{Nonadiabatic control of quantum transport fidelity in dissipative cold media}  

\author{Arnab Chakrabarti}
 \affiliation{Department of Chemical and Biological Physics, \\ Weizmann Institute of Science, Rehovot 7610001, Israel}
\affiliation{International Center of Quantum Artificial Intelligence for Science and Technology
(QuArtist) and Department of Physics, Shanghai University, 200444 Shanghai, China}
\author{Igor Mazets}
\affiliation{Vienna Center for Quantum Science and Technology (VCQ),\\ Atominstitut, TU Wien, 1020 Vienna, Austria}
\affiliation{Research Platform MMM “Mathematics–Magnetism–Materials”, c/o Fak. für Mathematik, Univ. Wien, 1090 Vienna, Austria}
\affiliation{Wolfgang Pauli Institute c/o Fak. für Mathematik, Univ. \\ Wien, Oskar-Morgenstern-Platz 1, 1090 Vienna, Austria}

\author{Tian-Niu Xu}
\affiliation{International Center of Quantum Artificial Intelligence for Science and Technology
(QuArtist) and Department of Physics, Shanghai University, 200444 Shanghai, China}

\author{Xi Chen}
\affiliation{Department of Physical Chemistry, University of the Basque Country UPV/EHU,\\ Apartado 644, 48080 Bilbao, Spain}

\author{Gershon Kurizki}
\affiliation{Department of Chemical and Biological Physics, \\ Weizmann Institute of Science, Rehovot 7610001, Israel}

\date{\today}

\begin{abstract}
We put forth a hitherto unexplored  control strategy that enables finite-speed, high-fidelity transport of a quantum wavepacket through a low-temperature dissipative medium. The control consists in confining the wavepacket  within a shallow anharmonic trap (tweezer), whose  nonuniform velocity is steered so as to maximize the transfer fidelity between two locations. A relevant scenario is a quantum impurity moving through an ultracold gas. Unlike shortcuts to adiabaticity, our approach can  simultaneously cope with wavepacket leakage via non-adiabatic and phonon-mediated  processes, provided both act perturbatively. Nor  does our approach require the application of compensating forces or counter-diabatic fields and thereby avoids the practical shortcomings of shortcut techniques.  Instead, optimal (highest fidelity) transport is achieved here by minimizing the functional overlap of the varying velocity-spectrum of the chosen trajectory with the combined (bath-induced and non-adiabatic) leakage spectrum.
\end{abstract}

\maketitle

\section{Introduction}

Transport of trapped atoms, ions or their collective excitations is an inherent part of quantum information processing protocols \cite{Rowe02}, quantum refrigeration cycles \cite{Rezek09} and   other useful tasks \cite{Leanhardt03, Miroschnychenko06}. Such transport should be desirably accomplished as fast as possible,  not only  to shorten the duty cycle of the process at hand, but also to minimize its inherently quantum hurdles: wavepacket spread and decoherence by the environment. Tweezers, i.e. moving field-induced traps, are commonly used to reduce the wavepacket leakage. However, unless the trap is very deep (which may require overly intense fields), fast trasport of the trapped atom triggers non-adiabatic transitions between the initially populated bound states of the trap and the continuum.  In an increasingly widespread approach, the non-adiabatic leakage is reduced by introducing a time-dependent transient, compensating ( alias  counter-diabatic) field) \cite{Torrontegui11} known as shortcut to adiabaticity.  Alternatively, if the trapping potential is so deep that it may be assumed harmonic, which is often not the case, an optimal  trajectory can be designed using Lewis-Riesenfeld invariants \cite{LR69, Chen11, Chen151, Chen152, Chen16}.  Yet, regardless of their advantages or drawbacks,the above methods cannot properly account for  the wavepacket decoherence  or leakage caused by its ubiquitous interaction  with the environment, alias a bath, since  they are based on a Hamiltonian description which fails for quantum dissipative processes. This conceptual gap is the motivation for the present work. It tackles the hitherto unexplored formidable problem  of finite-time (nonadiabatic) quantum transport in dissipative media by  a paradigmatic change of the universal dynamical control of  quantum system-bath interactions, previously introduced by  one of the authors  for discrete variables \cite{kurizki01}: here we construct  dynamical control that can effectively counter the quantum friction of moving wavepackets at minimal energy cost and maximal fidelity.

Specifically, we aim at developing a control strategy for achieving optimal (highest-fidelity) transport of a quantum wavepacket confined in a shallow, anharmonic potential trap,  by steering the trap motion through the environment, modelled as a finite-temperature quantum bosonic bath.   A central insight obtained from our approach is that  the coupling of a nonadiabatically    transported  wavepacket to the continuum is a   source of quantum friction that can be viewed as  an additional  effective bath (on top of the standard environmental bath), with acceleration-dependent coloured spectrum and temperature. Because of the inherent similarity of these baths,  we can express the loss probability at the end of the transport as the  overlap of the  time-dependent trap-speed spectrum ( the Fourier transform of its autocorrelation function) with the spectra of the non-adiabatic heat-bath, the environmental-bath and, quite unusually, their mutual cross-spectra. In order to maximize the survival probability, we find the optimal trajectory that minimizes these overlaps, and thus yields the highest transport fidelity  at the lowest energy cost,  as guaranteed by the Euler-Lagrange variational principle.

In order to facilitate analytic treatment of the solution, we assume a Morse- potential trap,  but the approach is suitable for any finite-depth potential. The wavepacket initially occupies the ground state of the trap, before starting to move  through the finite-temperature  bosonic medium.  We adopt the dressed Bose polaron \cite{Mazets05, Coalson19} description for the impurity wavepacket in the medium.  Its  controlled dynamics  is based  on a generalization of the Wigner-Weisskopf approach \cite{Coalson19}. Using the weak-coupling Fröhlich Hamiltonian \cite{Lampo19}, we restrict the dynamics to the single excitation sector of the BEC under the assumption of low temperature. We then calculate the survival probability in the bound state of the trap, while integrating out all other degrees of freedom and maximize it at the lowest energy cost by following  the Euler-Lagrange optimization, as was previously done by one of us for discrete variables \cite{kurizki10}.

\section{Model}

\begin{figure}[!h]
\centering
\hspace{0.5cm}\includegraphics[scale= 0.35]{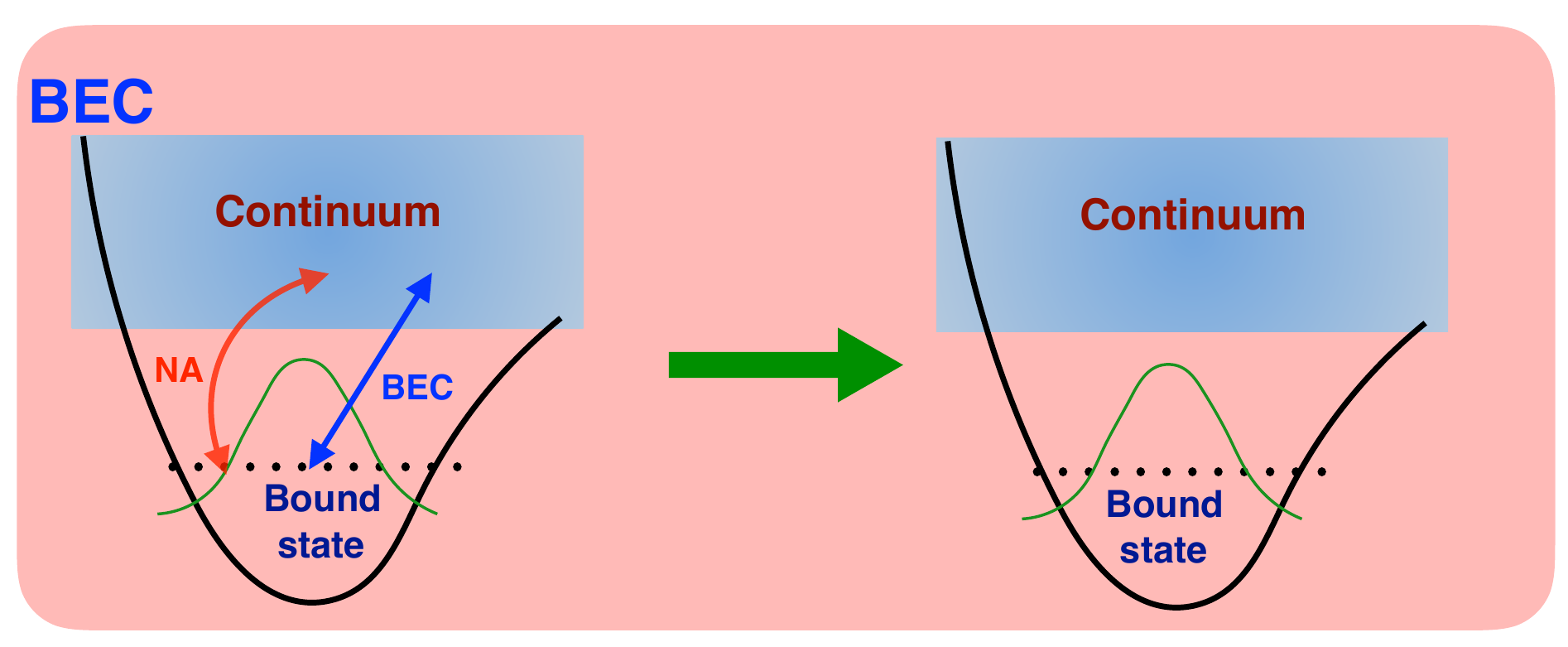}
\caption{Schematic diagram of the model under study}\label{S1}
\end{figure}

\noindent We consider an quantum impurity of mass $m$ in a moving trap (potential), immersed in a bath of interacting bosons of mass $m_B$ enclosed in a finite volume $\mathscr{V}$. The entire system (impurity + bosons) is described by the explicitly time-dependent Hamiltonian ($\hbar = 1$) \cite{Lampo19}

\begin{equation}\label{ham}
      H(t) = H_I(t) + H_B + H_{IB}(t),
\end{equation}
where,
\begin{equation}\label{hami}
       H_I(t) = \frac{p^2}{2m} + V[x - x_{\circ}(t)]
\end{equation}
\begin{equation}\label{hamb}
       H_B = \sum_{k \neq 0} \Omega_kb_k^{\dagger}b_k
\end{equation}
with $b^{\dagger}_k$ and $b_k$ defined as the creation and anihilation operators for a Bogoliubov phonon with momentum $k$, while $x$ and $p$ denote the position and conjugate momentum of the impurity. $x_{\circ}(t)$ denotes the time-dependent position of the centre of the trap. We shall choose the well-known Fr{\"o}lich model for the interaction Hamiltonian  \cite{Lampo19, Coalson19}

\begin{equation}\label{hamib2}
       H_{IB}(t) = \sum_{k\neq 0} V_k \, e^{ik\lbrace x - x_{\circ}(t)\rbrace}\, (b_k + b_{-k}^{\dagger}).
\end{equation}
For simplicity, we assume the moving trap $V[x - x_{\circ}(t)]$ to be shallow, such that it supports only a single bound-state along with a continuum of unbound (scattering) states. In our analysis, we restrict ourselves to a moving Morse-trap, but our treatment would apply to other potentials as well. Thus we choose

\begin{equation}\label{Morse}
       V[x - x_{\circ}(t)] = D\Big[ e^{-2a\lbrace x - x_{\circ}(t)\rbrace} - 2 e^{-a\lbrace x - x_{\circ}(t)\rbrace} \Big],
\end{equation}
where $D$ denotes the depth of the potential, $a$ is a parameter that assumes non-zero positive values determining the width of the Morse-trap, $x_{0}(t)$ is the time-dependent center of the well \cite{morse29}. A schematic diagram of our model is shown in Figure (\ref{S1})

We represent the Hamiltonian (\ref{ham}) in the instantaneous eigenbasis of the moving Morse potential, which admits both discrete and conitnuous energy eigen-states \cite{lima06, deffner15}. The discrete sector, enumerated by integer quantum numbers $n$, has discrete eigen-frequencies given by \cite{lima06, deffner15}

\begin{equation}\label{f1}
       \omega_n = -\frac{a^2}{2m}(N - n)^2,
\end{equation}
where $n$ ranges from 0 to the integer part of $N$, which is given by

\begin{equation}
       \Big(N + \frac{1}{2}\Big)^2 = \frac{2\,m\,D}{a^2}.
\end{equation}

\noindent The frequency dispersion relation for the continuous part is given by \cite{lima06, deffner15}

\begin{equation}\label{f2}
     \omega_{\kappa} = \frac{a^2}{2m}\kappa^2 \; , \; \; \forall \; \kappa \in [0, \infty],
\end{equation}
where $\kappa$ denotes the corresponding momentum of the unbound impurity state.

The transport causes the time-variation of the eigenfunctions of the Morse-potential, through their dependence on $[x - x_{\circ}(t)]$ (but the discrete and continuous eigenvalues are independent of the potential-center location). The explicit forms of the discrete and continuous eigen-functions $\lbrace \phi_n, \phi(\kappa) \rbrace$ are given in the Appendix \cite{lima06, deffner15}. We shall denote the instantaneous bound states by $\lbrace \vert n(t)\rangle \rbrace $ and the instantaneous continuum states by $\lbrace \vert \kappa (t)\rangle \rbrace$. Following \cite{kurizki01}, we shall represent the integrals over the continuum modes as summations, for simplicity.We shall account for the continuous character of the eigenvalues, whenever explicit evaluation of the integrals is necessary. After invoking the rotating-wave-approximation as in \cite{Kurizki94}, the Hamiltonian (\ref{ham}) in the instantaneous eigenbasis of the moving Morse-trap, takes the form \cite{zwiebach18}:

\begin{eqnarray}\label{ham3}
       H(t) & = & \omega_{\circ}\vert n(t)\rangle\langle n(t)\vert + \sum_{\kappa} \omega_{\kappa}\vert\kappa(t)\rangle\langle\kappa(t)\vert + \sum_{k \neq 0} \Omega_kb_k^{\dagger}b_k \nonumber\\
           &   & + \sum_{k \neq 0}\sum_{\kappa}\Big[ d_{n,\kappa}^k \vert n(t)\rangle\langle \kappa(t)\vert b_k^{\dagger}+ h.c.\Big] \nonumber\\
           &   & + \sum_{k \neq 0}\sum_{\kappa \leq \kappa'}\Big[ d_{\kappa,\kappa'}^k \vert \kappa(t)\rangle\langle \kappa'(t)\vert b_k^{\dagger} + h.c.\Big],
\end{eqnarray}
where the coefficients $d_{n,\kappa}^k = \langle n(t) \vert V_k\, e^{ik\lbrace x - x_{\circ}(t) \rbrace}\vert \kappa (t)\rangle$ and \\
$d_{\kappa,\kappa'}^k = \langle \kappa(t) \vert V_k\, e^{ik\lbrace x - x_{\circ}(t) \rbrace}\vert \kappa' (t)\rangle$ are independent of $x_{\circ}(t)$ and hence $t$, since the instantaneous Morse eigen-functions are expressed as functions of $\lbrace x - x_{\circ}(t)\rbrace$ [Appendix].

\section{Dynamics of the impurity}

Our goal is to maximize the survival probability of the impurity, in the bound state $\vert n (t)\rangle$ at the end of the transport process. We assume that the trapped impurity is immersed (quenched) at $t = 0$, into the ground state of the homogeneous medium. We shall adopt an extension of the Wigner-Weisskopf approach \cite{Coalson19} and as in \cite{mazets04}, we shall restrict our analysis to the single Bogoliubov excitation sector, since multiple phonon excitations are highly improbable in the limit of weak coupling and weak non-adiabaticity. The combined state of the impurity and the medium, at any time $t$ is then

\begin{eqnarray}\label{genstate}
  \vert\psi(t)\rangle & = & \alpha_{\circ}(t)e^{-i\omega_{\circ} t}\vert n(t); 0^B\rangle + \sum_{\kappa}\beta_{\kappa}(t)e^{-\omega_{\kappa} t}\vert \kappa (t) ; 0^B \rangle \nonumber\\
                      &  & + \sum_{k\neq 0} \alpha _{1k}(t) e^{-i(\omega_{\circ} + \Omega_k) t}\vert n(t) ; 1^B_k\rangle \nonumber\\
                      &  & + \sum_{k\neq 0} \sum_{\kappa}\beta _{\kappa 1 k}(t) e^{-i(\omega_{\kappa} + \Omega_k) t}\vert \kappa(t) ; 1^B_k\rangle,
\end{eqnarray}
while the initial state is
\begin{equation}\label{inistate}
       \vert\psi(0)\rangle = \vert n(0); 0^B\rangle.
\end{equation}
Here $\vert n(t); 0^B\rangle$ is the state with one impurity in the instantaneous bound state |n(t)i and the medium in the ground (vacuum) state. $\vert \kappa(t) ; 0^B\rangle$ indicates the impurity in the unbound (excited) state $\vert \kappa(t)\rangle$ while the BEC is still in the ground (vacuum) state. Similarly the states $\vert n(t); 1^B_k\rangle$ and $\vert \kappa(t); 1^B_k\rangle$ indicate the respective bound and unbound impurity states with a single Bogoliubov excitation in the medium, with momenum $k$. If we assume that initially the BEC is in the vacuum state with no Bogoliubov excitations, to the lowest order in $H_{IB}(t)$ we should have at most a single Bogoliubov phonon in the BEC, as is evident from (\ref{genstate}) \cite{Coalson19}. The non-adiabatic transitions due to the motion of the trap, cannot directly induce Bogoliubov excitations.

The time-evolution of the combined state $\vert\psi(t)\rangle$ is governed by the Schr{\"o}dinger equation

\begin{equation}\label{SE}
          i\frac{d}{dt}\vert \psi(t)\rangle = H(t)\vert \psi(t)\rangle.
\end{equation}
Using (\ref{ham3}), (\ref{genstate}) in (\ref{SE}) we take inner-products on both sides with respect to $\langle n(t); 0^B\vert$ , $\langle \kappa_1(t); 0^B\vert$ , $\langle n(t); 1^B_{k_1}\vert$ and $\langle \kappa_1(t) ; 1^B_{k_1}\vert$ to get the following dynamical equations for the arbitary coefficient vectors, \begin{equation}\label{Bvector}
      \bm{B} = 
\begin{bmatrix}
\beta_{\kappa}  \\
\beta_{\kappa 1 k} \\
\end{bmatrix}
\end{equation}
and
\begin{equation}
\bm{A} = 
\begin{bmatrix}
\alpha_{\circ}  \\
\alpha_{1k} \\
\end{bmatrix}
\end{equation}
as
\begin{eqnarray}\label{dyn4}
       \frac{d}{dt}\bm{A} & = & -i\bm{F}^{\dagger}\bm{B}\nonumber\\
       \frac{d}{dt}\bm{B} & = & \bm{M}\bm{B} - i\bm{F}\bm{A},
\end{eqnarray}
where $\bm{M} = \begin{bmatrix}
-\gamma_{\kappa\kappa'} & -id_{\kappa\kappa'}^{k*}  \\
-id_{\kappa\kappa'}^k & - \gamma_{\kappa\kappa'} \\
\end{bmatrix}$ and $\bm{F} = \begin{bmatrix}
i\gamma_{n\kappa}^{*} & d_{n\kappa}^{k*}  \\
0 & i\gamma_{n\kappa}^{*} \\
\end{bmatrix}$.  
\\\\
\noindent In the above expression
\begin{eqnarray}\label{nonad}
       \gamma_{ab}(t) = e^{-i\omega_{ba}t}\frac{\langle a(t)\vert \dot{H}_I(t)\vert b(t)\rangle}{(\omega_{b} - \omega_a)} \, ; \, \omega_{ba} = \omega_{b} - \omega_{a}\\
       \, ;\forall \, a,b \in \lbrace n,\kappa\rbrace \nonumber
\end{eqnarray}
denotes the non-adiabatic transition matrix-elements
and
\begin{eqnarray}\label{couple}
       d_{ab}^k(t) = e^{-i(\omega_{ba} - \Omega_k)t} d_{ab}^k\, ; \, \omega_{ba} = \omega_{b} - \omega_{a} \\
       \, ;\,\forall \; a,b \in \lbrace n,\kappa\rbrace \nonumber
\end{eqnarray}
are the dipole-transition matrix elements. We can write the general expression for bound-bound and bound-continuum non-adiabatic transition rates using
\begin{equation}\label{ad3}
      \langle n(t) \vert \frac{d}{dt}\vert k(t)\rangle = \frac{\langle n(t)\vert\dot{H}_{\rm I}(t)\vert \kappa(t)\rangle}{\omega_{\kappa n}} = \dot{x}_{\circ}(t)\mu_{n\kappa},
\end{equation}
where $\mu_{n\kappa} = \frac{\widetilde{\mu}_{n\kappa}}{\omega_{\kappa n}}$ and $\widetilde{\mu}_{n\kappa} = \int\limits_{-\infty}^{\infty} dq \, \phi_n(q)\Big[ 2\,a\,D\lbrace e^{-2aq} - e^{-aq}\rbrace\Big]\phi_{\kappa}(q)$ is independent of $t$. We cannot directly extend equation (\ref{ad3}) to describe continuum-continuum non-adiabatic transitions, since it it tends to diverge, suffering from ``the problem of small denominators'' \cite{polkovnikov17}. The problem may, in principle, be avoided by introducing a ``virtual gap'' in the continuous spectrum through re-defining the continuum eigen-functions $\phi[\kappa,z(t)]$ as We\'yl eigen-differential wave packets, which behave as if they were discrete eigen-functions \cite{maamacheprl08, maamachepra08}. One can then arrive at the limiting (finite) value of the non-adiabatic transition rates for a gapless spectrum \cite{maamacheprl08, maamachepra08}. However, this is not required for the time-scales we wish to explore.

\begin{figure}[!h]
\centering
\hspace{0.5cm}\includegraphics[scale= 0.35]{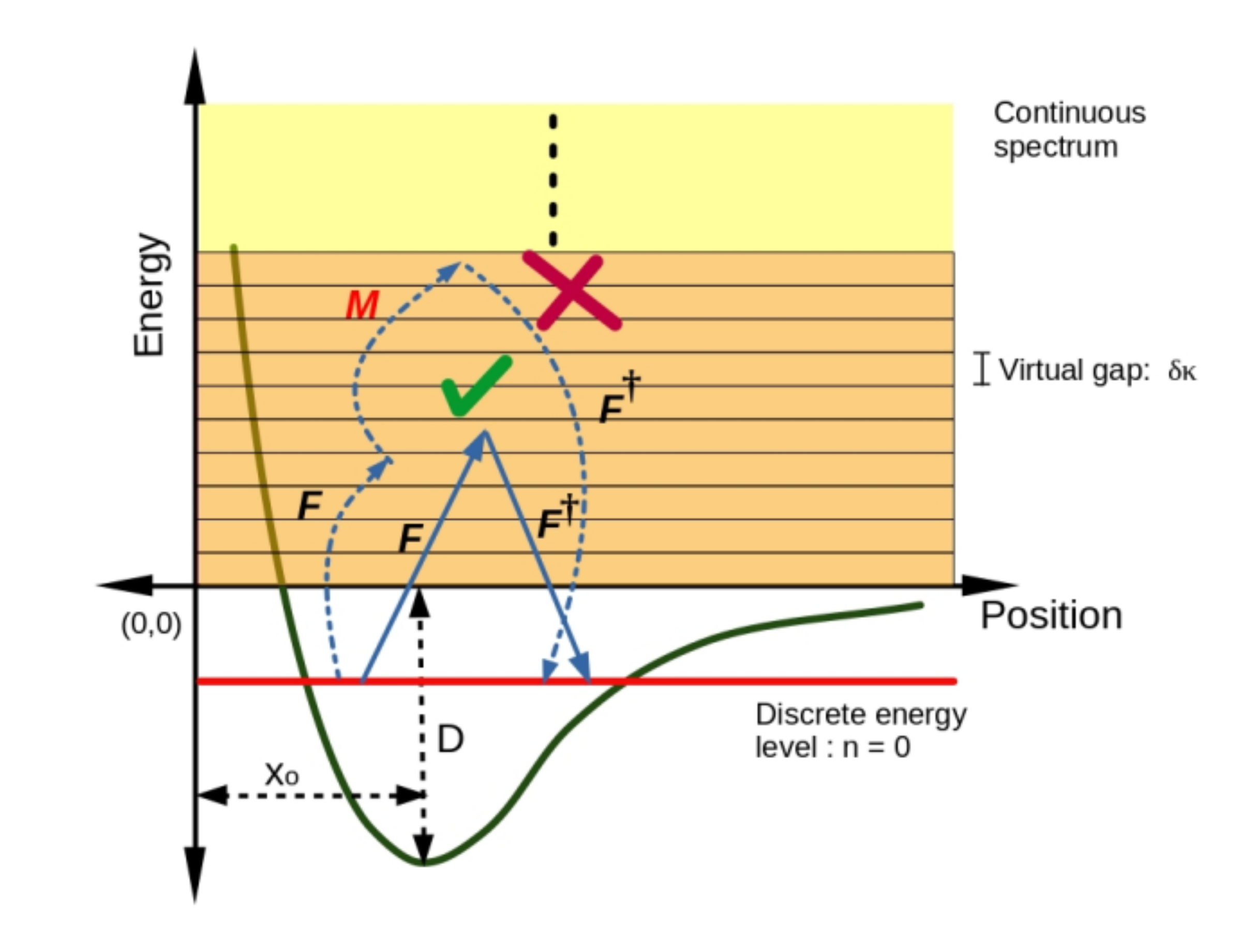}
\caption{Schematic representation of the Short-time dynamics}\label{S2}
\end{figure}

In order to find an expression for the survival probability in the bound-state, we formally solve the second equation in (\ref{dyn4}) to have

\begin{eqnarray}\label{sol1}
     \bm{B}(t) & = & T\,e^{-\int\limits_0^t\,ds \bm{M}(s)}\bm{B}(0) \nonumber\\
     &  & -\; i\int\limits_0^t \; ds\; T\;e^{-\int\limits_s^t\,d\tau \bm{M}(\tau)}\,\bm{F}(s)\;\bm{A}(s), 
\end{eqnarray}
where $T$ denotes the chronological time-ordering operator. Since we assume that initially the impurity was trapped in the Morse potential, we must have $\bm{B}(0) = \bm{0}$. Then equation (\ref{sol1}) reduces to

\begin{eqnarray}\label{sol2}
     \bm{B}(t)  =  -\;i\int\limits_0^t \, ds\; \bm{U_M}(t,s)\;\bm{F}(s)\;\bm{A}(s),
\end{eqnarray}
where we have defined $\bm{U_M}(t,s) = T\,\exp[-\int\limits_s^t\,d\tau \bm{M}(\tau)]$. Substituting (\ref{sol2}) on the r.h.s. of the first equation in (\ref{dyn4}), we then have

\begin{eqnarray}\label{dyn5}
       \frac{d}{dt}\bm{A} & = & -\int\limits_0^t \, ds\, \Big[\bm{F}^{\dagger}(t)\bm{U_M}(t,s)\bm{F}(s)\Big]\,\bm{A}(s)
\end{eqnarray}
Equation (\ref{dyn5}) cannot be generally solved without further approximations. To this end, we assume that the time-interval of the transport $(t - 0)$ is much smaller than the time-scales in which both $\gamma_{ab}(t)$ and $d_{ab}^k(t)$ change appreciably. In such a scenario, we can approximate (\ref{dyn5}) to second-order in $\dot{x}_{\circ}(t)$ and $\vert\vert V_k\vert\vert$. We do this by replacing $\bm{U_M}(t,s)$ by an identity operator and $\bm{A}(s)$ by $\bm{A}(t)$ in the kernel of (\ref{dyn5}). The physical meaning of this approximation is schematically expressed in Figure (\ref{S2})

We thus have

\begin{equation}\label{dyn6}
     \frac{d}{dt}\bm{A}   =  -\int\limits_0^t \, ds\,\bm{F}^{\dagger}(t)\bm{F}(s)\,\bm{A}.
\end{equation}
Integrating (\ref{dyn6}) and retaining the lowest-order term in $\bm{F}$ we then have

\begin{equation}\label{sol3}
     \bm{A}(t) = \bm{A}(0) - \int\limits_0^t\, dt_1\int\limits_0^{t_1} \, dt_2\,\bm{F}^{\dagger}(t_1)\bm{F}(t_2)\,\bm{A}(0) + \mathcal{O}[\bm{F}^4].
\end{equation}
The survival probability of the bound-state $\vert n(t)\rangle$ is then given by 

\begin{eqnarray}\label{Sprob}
  P_n(t) & = & \vert \bm{A}(t) \vert^2 \nonumber\\
         & = & \Big\vert 1 - \int\limits_0^t\, dt_1\int\limits_0^{t_1}\,dt_2\,\Gamma_{n}(t_1, t_2) \Big\vert^2 \nonumber\\
         &   &  + \sum_{k \neq 0} \Big\vert -i\int\limits_0^t\, dt_1\int\limits_0^{t_1}\, dt_2\; C_n^k(t_1,t_2) \Big\vert^2,
\end{eqnarray}
where $\Gamma_{n}(t_1, t_2) = \gamma_{n\kappa}(t_1)\gamma_{n\kappa}^{*}(t_2)$ , $C_n^k(t_1,t_2) = d_{n\kappa}^{k}(t_1)\gamma_{n\kappa}^{*}(t_2) $.

We note that each term in the above expression is non-negative and real-valued. The first term on the r.h.s. of (\ref{Sprob}) represents the contribution arising solely from non-adiabatic transitions, while the second term represents a cross-correlation between non-adiabatic and impurity-phonon interaction processes. In the absence of coupling to the medium ( $\vert\vert V_k \vert\vert = 0$), the survival probability $P_n^{\rm free}(t)$ would depend only on the non-adiabatic transitions, i.e. 

\begin{equation}\label{Sprobfree}
    P_n^{\rm free}(t) = \Big\vert 1 - \int\limits_0^t\, dt_1\int\limits_0^{t_1}\,dt_2\,\Gamma_{n}(t_1, t_2) \Big\vert^2.
\end{equation}
Thus, we have 
\begin{eqnarray}
   P_n(t) - P_n^{\rm free}(t) & = & \int\limits_0^{\infty}d\omega\,\rho(\omega) \Big\vert -i\int\limits_0^t\, dt_1\int\limits_0^{t_1}\, dt_2 \; C_n(\omega)(t_1,t_2) \Big\vert^2  \nonumber\\
                              &   & \geq 0.
\end{eqnarray}

\noindent Remarkably and counter-intuitively, equation (\ref{Sprob}) shows that on a suitably short-time scale, dissipative coupling to the medium is advantageous, resulting in higher survival probabilities of the bound-state of the moving impurity. This occurs due to the joint effect of the non-adiabatic and dissipative transition processes, expressed through their cross-term $C_n^k(t_1,t_2)$. Yet, if the total duration of the transport process becomes too long, the survival probability may decrease when phonon-mediated relaxation becomes significant. 

An exact solution of (\ref{dyn6}) is complicated, involving higher-order braided interaction terms between the nonadiabatic and impurity-phonon coupling effects. Accurate analysis of this equation  is beyond the scope of a systematic perturbation theory,  typically above the domain of applicability of the  first few  perturbation orders. A full numerical treatment  is then called for. Nonetheless, the short-time solution presented above, clearly shows that survival probability can in principle be enhanced by an otherwise dissipative, phonon emission process(!).


\section{Optimal Control}
\noindent Our aim is to find the optimal trajectory which maximizes the survival probability $P_n(t)$ at the end of the transport process. To this end we first consider (\ref{Sprobfree}), which forms the first term on the r.h.s. of (\ref{Sprob}). 

\begin{eqnarray}\label{Sprobfree1}
      P_n^{\rm free}(t) & = & \Big[1 - \int\limits_0^t\, dt_1\int\limits_0^{t_1}\,dt_2\,\dot{x}_{\circ}(t_1)\dot{x}_{\circ}(t_2)\,\Phi(t_1 - t_2)\Big]\nonumber\\
      &  & \Big[1 - \int\limits_0^t\, dt_1'\int\limits_0^{t_1'}\,dt_2'\,\dot{x}_{\circ}(t_1')\dot{x}_{\circ}(t_2')\,\Phi(t_1' - t_2')\Big]^*, 
\end{eqnarray}
where we have used (\ref{nonad} and \ref{ad3}), while defining the memory kernel
\begin{equation}\label{nonadcorr}
    \Phi(t) = \sum_{\kappa} e^{-i \omega_{\kappa n} t} \frac{\vert \widetilde{\mu}_{n \kappa} \vert^2}{\omega_{\kappa n}^2}.
\end{equation}
Simplifying, we have
 
\begin{eqnarray}\label{Sprobfree3}
   P_n^{\rm free}(t) & = & 1 - \int\limits_0^t\, dt_1\int\limits_0^{t_1}\,dt_2\,\dot{x}_{\circ}(t_1)\dot{x}_{\circ}(t_2)\Big\lbrace 2\text{Re}[\Phi(t_1 - t_2)]\Big\rbrace \nonumber\\
                     &   &  \; + \;\mathscr{O}[ (\dot{x}_{\circ})^4] \,;
\end{eqnarray}
 the last term on the r.h.s. of the above equation is of the order of $(\dot{x}_{\circ})^4$ and hence negligible with respect to the other terms in the expression. 
 
\begin{figure}[!h]
\centering
\hspace{1.5cm}\includegraphics[scale= 0.7]{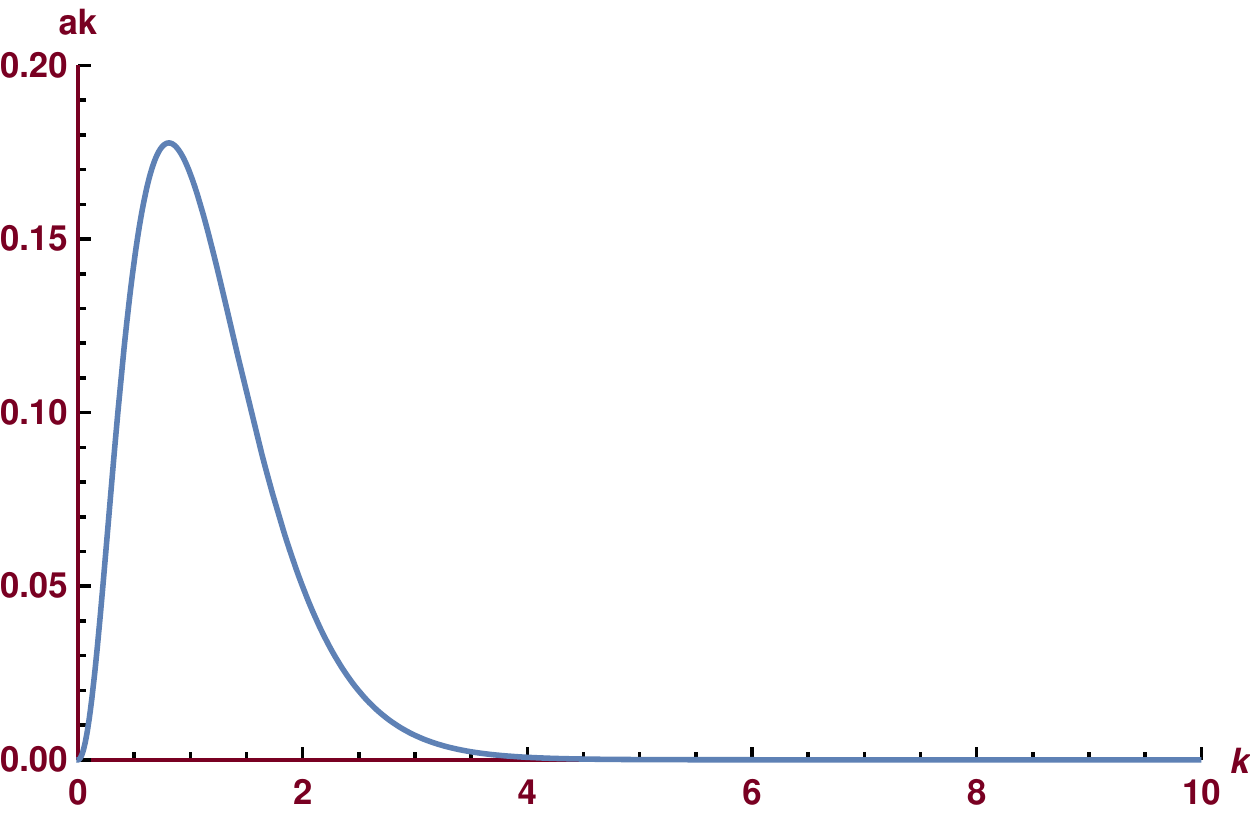}
\caption{Dependence of $a_{\kappa} = \vert \widetilde{\mu}_{0\kappa} \vert^2/\omega_{\kappa 0}^2$ on $\kappa$.}\label{F1}
\end{figure}

Defining $\mathcal{G}(\omega) = \int\limits_{-\infty}^{\infty} dt\, e^{i \omega t}\,2\,\text{Re}[\Phi(t)]$, $p_{\circ}(t) = \dot{x}_{\circ}(t)$ and $p_{\circ}(t,\omega) = \int\limits_0^t dt_1\, e^{-i\omega t_1} p_{\circ}(t_1)$ while neglecting the $4^{\rm{th}}$ order contributions, we can then write, 

\begin{equation}\label{SprobKK}
     P_n^{\rm free}(t) = 1 - \frac{1}{2\pi}\int\limits_{-\infty}^{\infty} d\omega \,\mathcal{G}(\omega)\big\vert p_{\circ}(t,\omega)\big\vert^2 ,
\end{equation}
which obeys the universal formula of Kofman and Kurizki \cite{kurizki01}. We need to minimize

\begin{equation}\label{J1}
      J_1 = \frac{1}{2\pi}\int\limits_{-\infty}^{\infty} d\omega \,\mathcal{G}(\omega)\big\vert p_{\circ}(t,\omega)\big\vert^2
\end{equation}
in order to obtain the maximum survival probability $P_n^{\rm {free} }(t)$. To achieve this, we need to evaluate $\mathcal{G}(\omega)$ and $\text{Re}[\Phi(t)]$ [see Appendix]. Defining $a_{\kappa} = \vert \widetilde{\mu}_{0\kappa} \vert^2/\omega_{\kappa 0}^2$ we can write 

\begin{equation}\label{nonadcorr2}
\text{Re}[\Phi(t)] = \sum_{\kappa} a_{\kappa}\,\cos(\omega_{\kappa 0} t).
\end{equation}
The r.h.s. of the above equation is a weighted sum (integral) of cosine functions and can, in principle, assume both positive and negative real values for different $t$ -- in contrast to ordinary correlation functions describing damping phenomena.
Assuming $ a = 1 $, $ m = 1$ and $D = \frac{1}{2}$ for our problem, we have $N = \frac{1}{2}$ which signifies that the Morse potential only supports a single bound state $(n = 0)$ as required in our case. One can then numerically estimate the coefficients $a_{\kappa}$ as a function of $\kappa$. The plot of $a_{\kappa}$ versus $\kappa$ is shown in Figure \ref{F1}.

\begin{figure}[!h]
\centering
\hspace{0.5cm}\includegraphics[scale= 0.6]{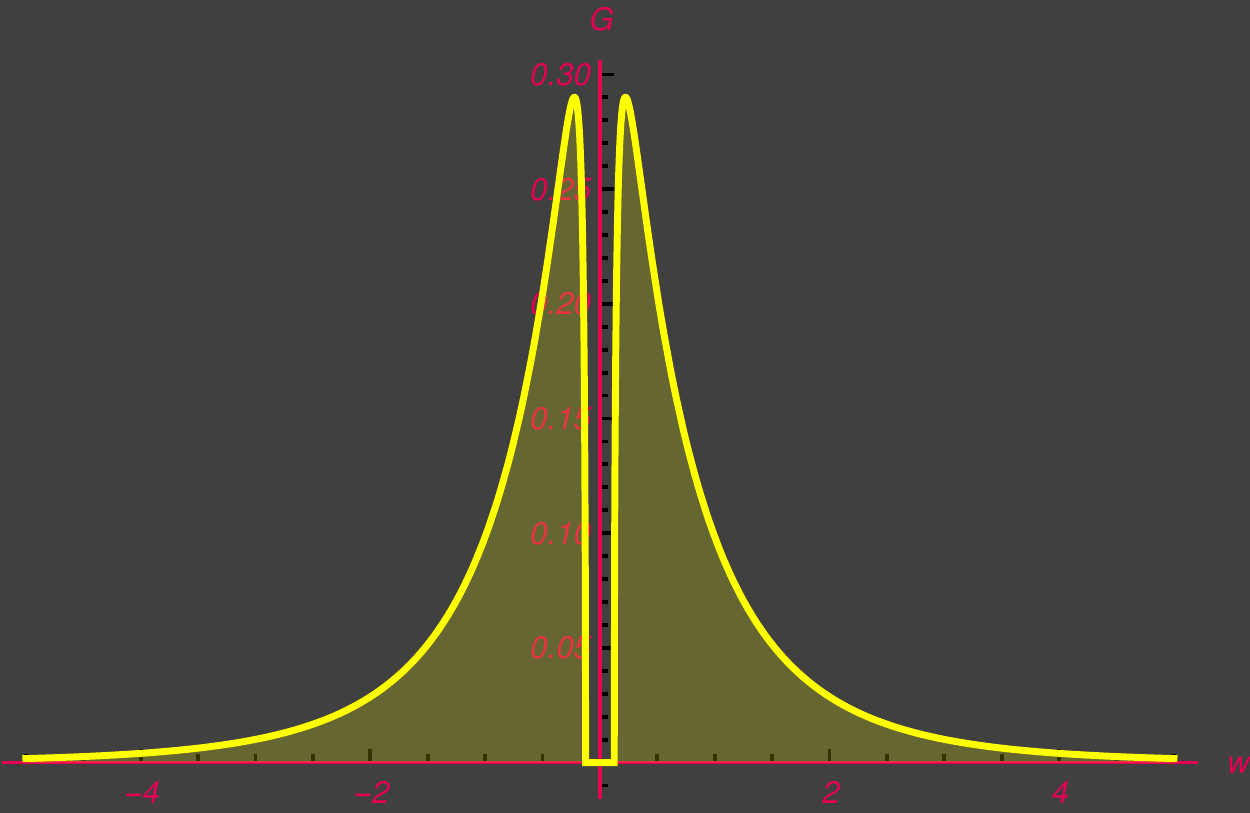}
\caption{Spectrum of non-adiabatic transitions, $\mathcal{G}(\omega)$ versus $\omega$.}\label{F2}
\end{figure}

The spectrum of the non-adiabatic couplings is given by 

\begin{eqnarray}\label{spect1}
      \mathcal{G}(\omega) & = & \int\limits_{-\infty}^{\infty} dt\, e^{i \omega t}\;\sum_{\kappa} a_{\kappa} [\;e^{i\omega_{\kappa 0} t} + e^{-i\omega_{\kappa 0} t}\;] \nonumber\\
                           & = & \sum_{\kappa} a_{\kappa}\;[\;\delta(\omega + \omega_{\kappa 0}) + \delta(\omega - \omega_{\kappa 0})\;]
\end{eqnarray}
We shall rewrite the discrete summation over $\kappa$ in the form of a continuous line integral over $\mathbb{R}_{\geq}$ and express $a_{\kappa}$ as $a(\kappa)$. We can then explicitly evaluate $\mathcal{G}(\omega)$ is a piece-wise continuous function in $\omega$ having the explicit form

\begin{equation*}\label{nonadspect}
\mathcal{G}(\omega) = \begin{cases}
 \sqrt{\frac{m^*}{2}} \;\frac{a[\kappa_2(\omega)]}{\sqrt{\omega_0 + \omega}}& \omega > -\omega_0\\
0 & \omega_0 \leq \omega \leq -\omega_0\\
\sqrt{\frac{m^*}{2}} \;\frac{a[\kappa_1(\omega)]}{\sqrt{\omega_0 - \omega}} & \omega < \omega_0,
\end{cases}
\end{equation*}
where $\kappa_1(\omega) = \sqrt{2m^*(\omega_0 - \omega)}$ and $\kappa_2(\omega) = \sqrt{2m^*(\omega_0 + \omega)}$. The spectrum of the non-adiabatic transitions, $\mathcal{G}(\omega)$ is shown in Figure (\ref{F2}).

\subsection{Euler-Lagrange Optimization}
\begin{figure}[!h]
\centering
\hspace{0.5cm}\includegraphics[scale= 0.55]{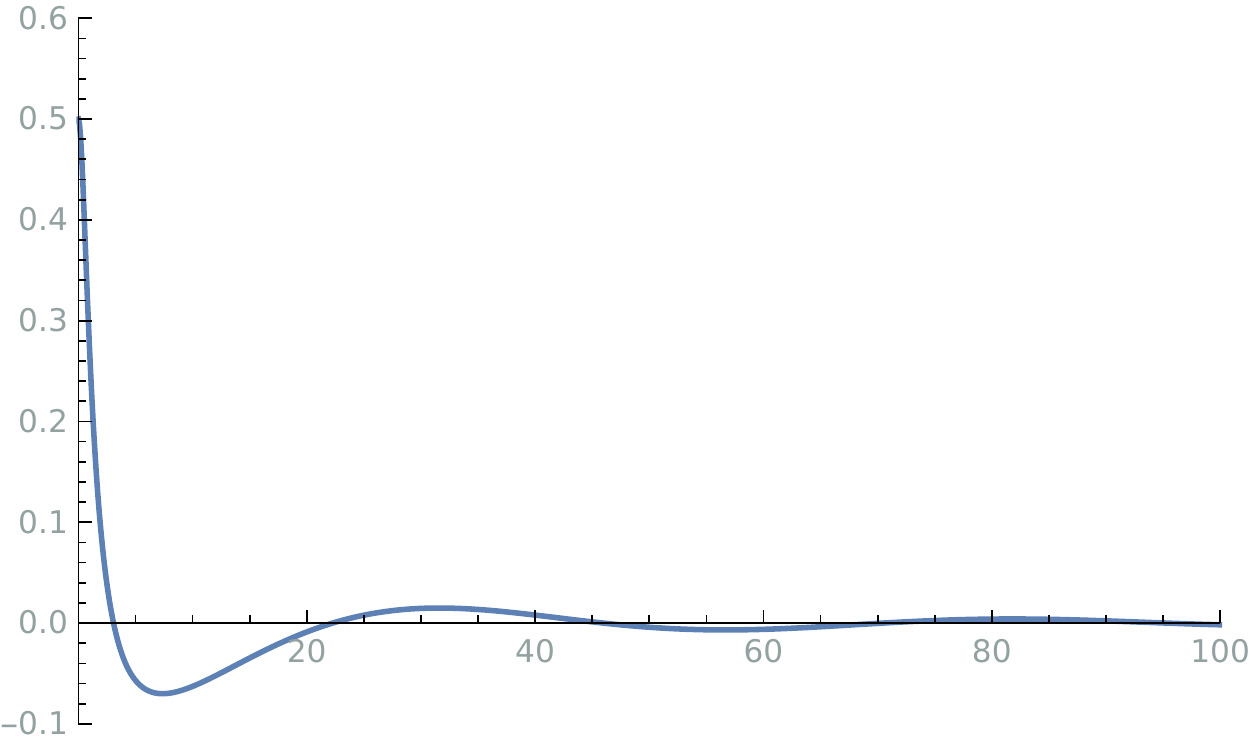}
\caption{Memory kernel, $2 {\rm Re}[\Phi(t)] $ as a function of  $t$.}\label{F3}
\end{figure}

We need to find the optimal function $p_{\circ}(t)$ that minimizes $J_1$. In time-domain, $J_1$ is given by

\begin{equation}\label{J11}
       J_1[p_{\circ}] = \int\limits_0^t\, dt_1\int\limits_0^{t}\,dt_2\,p_{\circ}(t_1)p_{\circ}(t_2)\;\Big\lbrace 2\,\text{Re}[\Phi(t_1 - t_2)]\Big\rbrace.
\end{equation}
In order to obtain non-trivial and physically meaningful results for the control functional, $p_{\circ}(t)$, following \cite{kurizki12}, we shall impose a constraint of the form

\begin{equation}\label{constraint1}
       \int\limits_0^t \, d\tau\, [\dot{p}_{\circ}(\tau)]^2 = E,
\end{equation}
where $E$ is constant, indicating the fluence of the variational parameter $p_{\circ}$. The resulting Euler-Lagrange equation is,

\begin{equation}\label{ELeq}
    \lambda\,\ddot{p}_{\circ}(t) = \int\limits_0^t p_{\circ}(\tau)\;\Big\lbrace 2\,\text{Re}[\Phi(t - \tau)]\Big\rbrace,
\end{equation}
where $\lambda$ is the Lagrange multiplier. Note that $\lambda$ can assume positive and negative real values as shown in Appendix. The integro-differential equation (\ref{ELeq}) can be solved in terms of the Laplace transforms:

\begin{eqnarray}\label{Laplace1}
   G(s) & = & \mathcal{L}\Big[2\,\text{Re}[\Phi(t)]\,;\,s\Big]\\
   P_{\circ}(s) & = & \mathcal{L}\Big[p_{\circ}(t)\,;\,s\Big].
\end{eqnarray}
We can represent the Euler-Lagrange equation (\ref{ELeq}) in the Laplace domain as

\begin{equation}\label{ELLaplace}
   \lambda \Big[\; s^2\,P_{\circ}(s) - s\,p_{\circ}(0) - \dot{p}_{\circ}(0)\;\Big] = P_{\circ}(s)\,G(s),
\end{equation}
Assuming that trap is initially at rest i.e. $\dot{p}_{\circ}(t) = 0$, we can then solve (\ref{ELLaplace}) as

\begin{eqnarray}\label{solution1}
     P_{\circ}(s) = \frac{\dot{p}_{\circ}(0)}{\Big[\,s^2 - \frac{1}{\lambda}\,G(s)\,\Big]}.
\end{eqnarray}
It is evident from equation (\ref{solution1}) that, in order to have a non-trivial solution for $p_{\circ}(t)$, initial acceleration of the trap, $\dot{p}_{\circ}(0) \neq 0$. This serves as a ``sanity check" since we must have non-zero initial acceleration in order to effect a transport. An exact calculation of $G(s)$ from (\ref{nonadcorr2}) is difficult since it involves an integral over all $\kappa$ modes. But we can numerically estimate $2{\rm Re}[\Phi(t)]$, as

\begin{equation}
    2 {\rm Re}\Phi(t) = \int\limits_0^{\infty} \, d\kappa \; \; 2a_{\kappa} \cos[\omega_{\kappa 0}(\kappa) t] 
\end{equation}
and plot it as a function of time, as shown in Figure (\ref{F3}).

\begin{figure}[!h]
\centering
\hspace{0.5cm}\includegraphics[scale= 0.55]{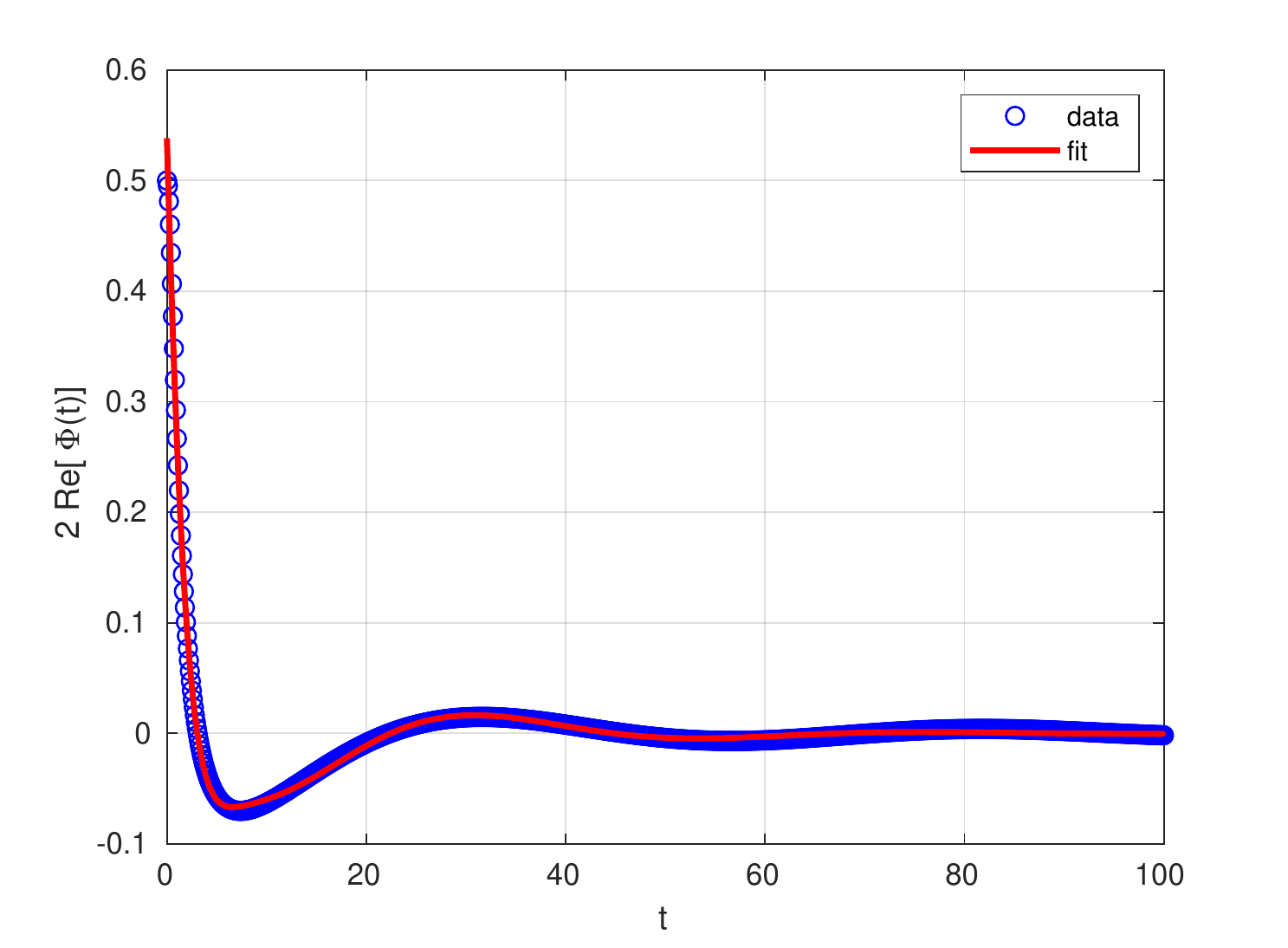}
\caption{Fitting of the memory kernel, $2 {\rm Re}[\Phi(t)] $ with the model $g(t)$.}\label{F4}
\end{figure}

Figure (\ref{F3}) shows that the memory kernel has a Non-Markovian behaviour, which is akin to the behaviour of a damped oscillator. Consequently, we approximate this with a general exponentially decaying oscillatory function of the form

\begin{equation}
        g(t) = \sum_j\Big[a_j\,e^{-b_j\,t}\,\cos(w_{1j} t) + c_j\,e^{-d_j\,t}\,\sin(w_{2j} t)\Big] \, u(t).
\end{equation}
where $u(t)$ denotes the unit step function. Using $g(t)$ to fit the data in Figure. {F3}, we find that the fitting function is of the form 
\begin{equation}
        g(t) = \Big[a_1\,e^{-b_1\,t}\,\cos(w_1 t) + c_1\,e^{-_1d\,t}\,\sin(w_2 t)\Big] \, u(t),
\end{equation}
with $a_1 = 0.5383 \, , b = 0.5831\, , c = - 0.1054, \, d = 0.0576, \, w_1 = - 0.3782, \, w_2 = 0.14$. The fit of $g(t)$ with the data in Figure (\ref{F3}) is shown in Figure. (\ref{F4}).

\begin{figure}[!h]
\centering
\hspace{0.5cm}\includegraphics[scale= 0.55]{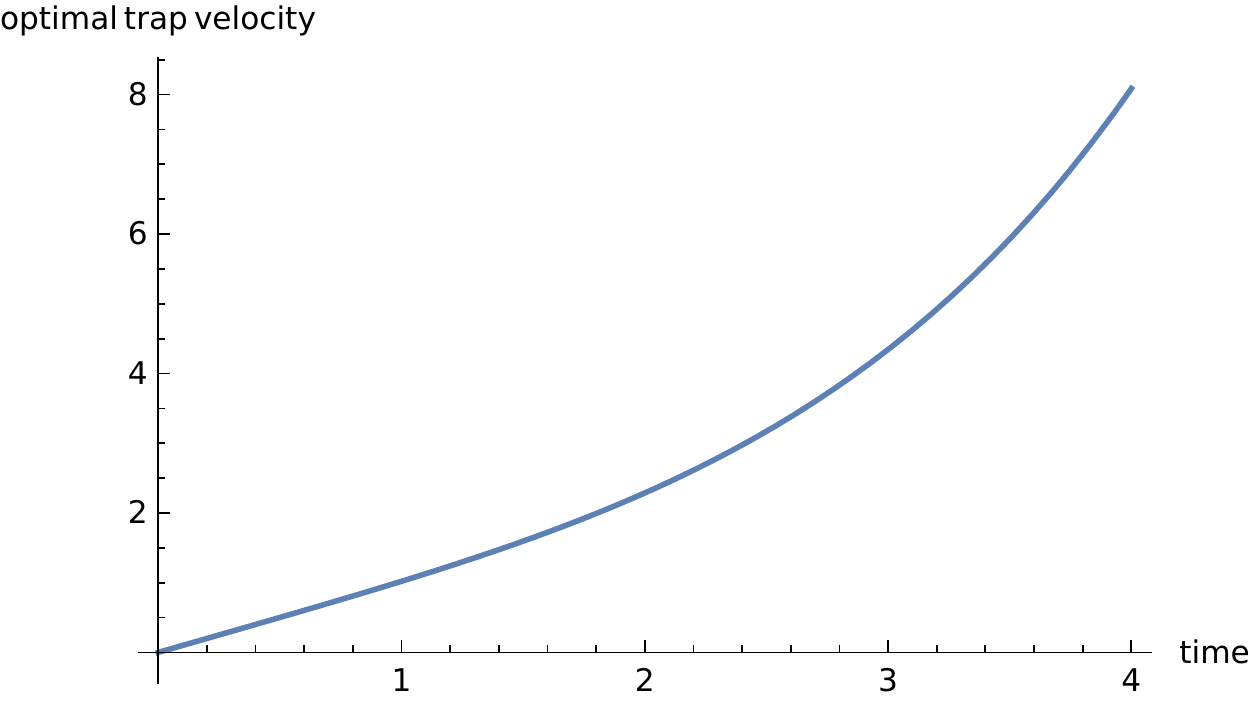}
\caption{Optimal $p_{\circ}(t)$ as a function of $t$ for $\lambda = 1$.}\label{F5}
\end{figure}

\noindent We can then use the approximation $2 {\rm Re}[\Phi(t)] \approx g(t)$, to find $G(s)$ using (\ref{Laplace1}). In order to find the solution $p_{\circ}(t)$ from (\ref{solution1}), we then have to simply find the poles of $s^2 - \frac{1}{\lambda}\,G(s)$ and calculate residues at these poles. For simplicity, we shall assume $\dot{p}_{\circ}(0) = 1$ for our numerical estimations. Through explicit numerical calculations, we find that in general, $s^2 - \frac{1}{\lambda}\,G(s)$ has $6$ distinct simple poles which depend on the value of the Lagrange multiplier $\lambda$. For $\lambda > 0 $ there is atleast one real pole, $\alpha > 0$, which results in a rapidly diverging exponential behaviour of $p_{\circ}(t)$ as shown in Figure (\ref{F5}). Such diverging exponential solutions are of little practical importance.

\begin{figure}[!h]
\centering
\hspace{0.5cm}\includegraphics[scale= 0.6]{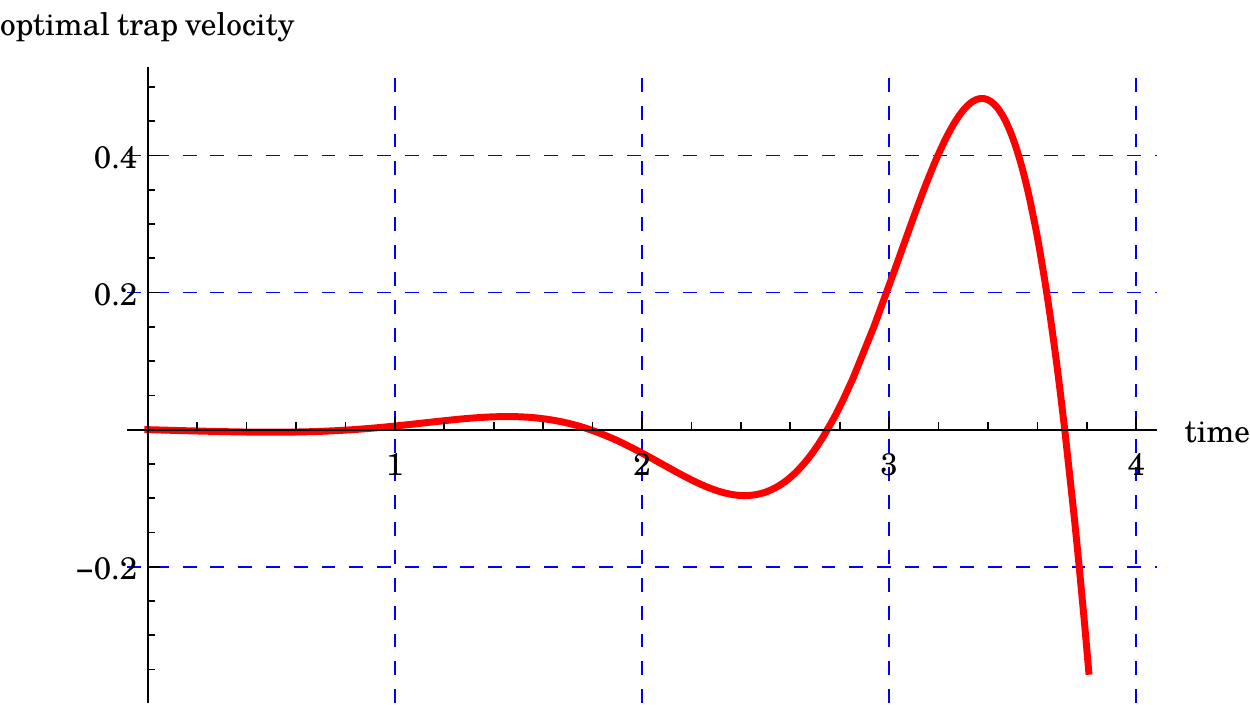}
\caption{Optimal $p_{\circ}(t)$ as a function of $t$ for $\lambda = -0.01$.}\label{F6}
\end{figure}

\noindent On the other hand for $\lambda < 0$, all poles with $\rm{Re}(s) > 0$ appear as complex conjugate pairs, resulting in an growing oscillatory behaviour. The corresponding solutions, offer a reliable choice for the optimal trajectory, as long as they predict physically realizable values for the survival probability. At large times, owing to the poles on the right hand side of the imaginary axis, the solutions eventually diverge, indicating the break-down of our original second-order approximation. For $\lambda < 0$, we further find that, we can have longer lifetimes in the bound state with decreasing magnitude of $\lambda$. The optimal trajectory for $\lambda = -0.01$ is shown in Figure (\ref{F6})

\begin{figure}[!h]
\centering
\hspace{0.5cm}\includegraphics[scale= 0.6]{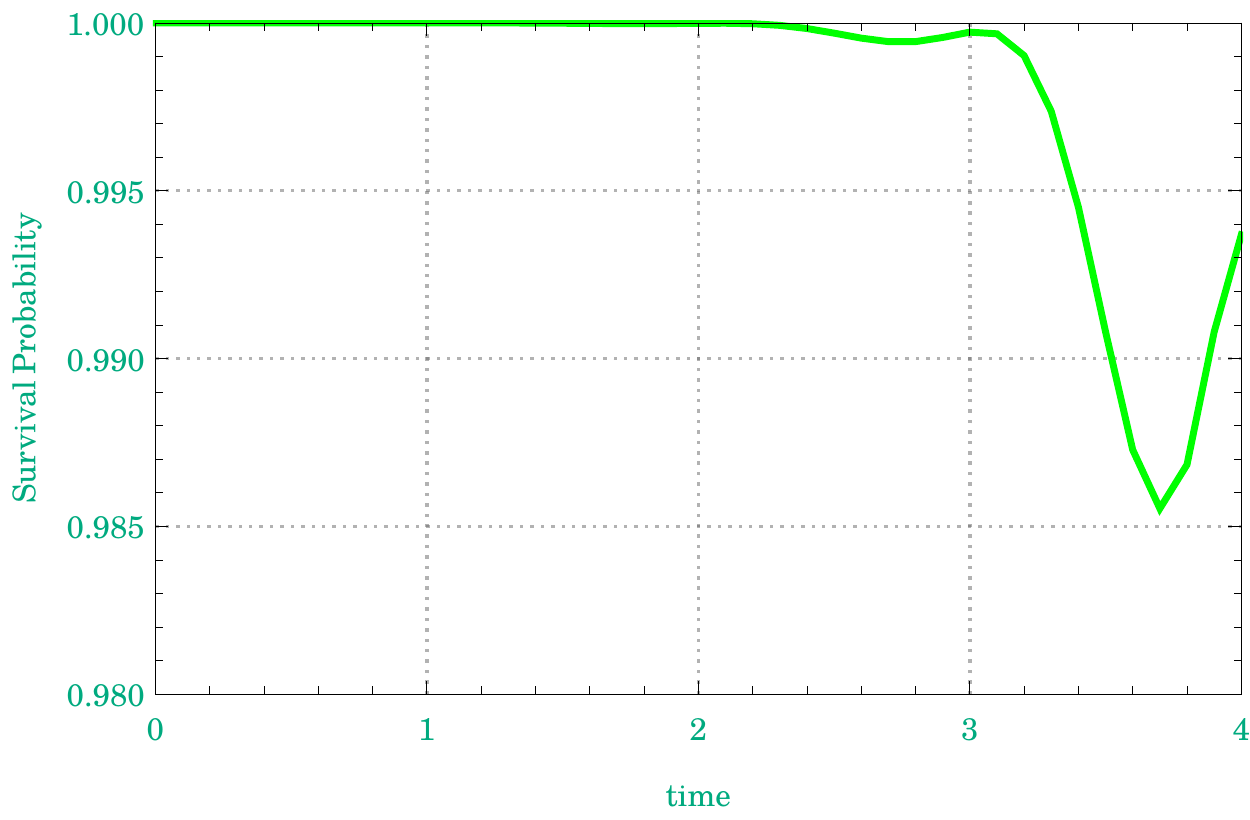}
\caption{Survival probability along the optimal trajectory $p_{\circ}(t)$ as a function of $t$ for $\lambda = -0.01$.}\label{F7}
\end{figure}

\noindent and the corresponding time-dependence of the survival probability in the bound state is shown in Figure (\ref{F7}).

\begin{figure}[!h]
\centering
\hspace{0.5cm}\includegraphics[scale= 0.6]{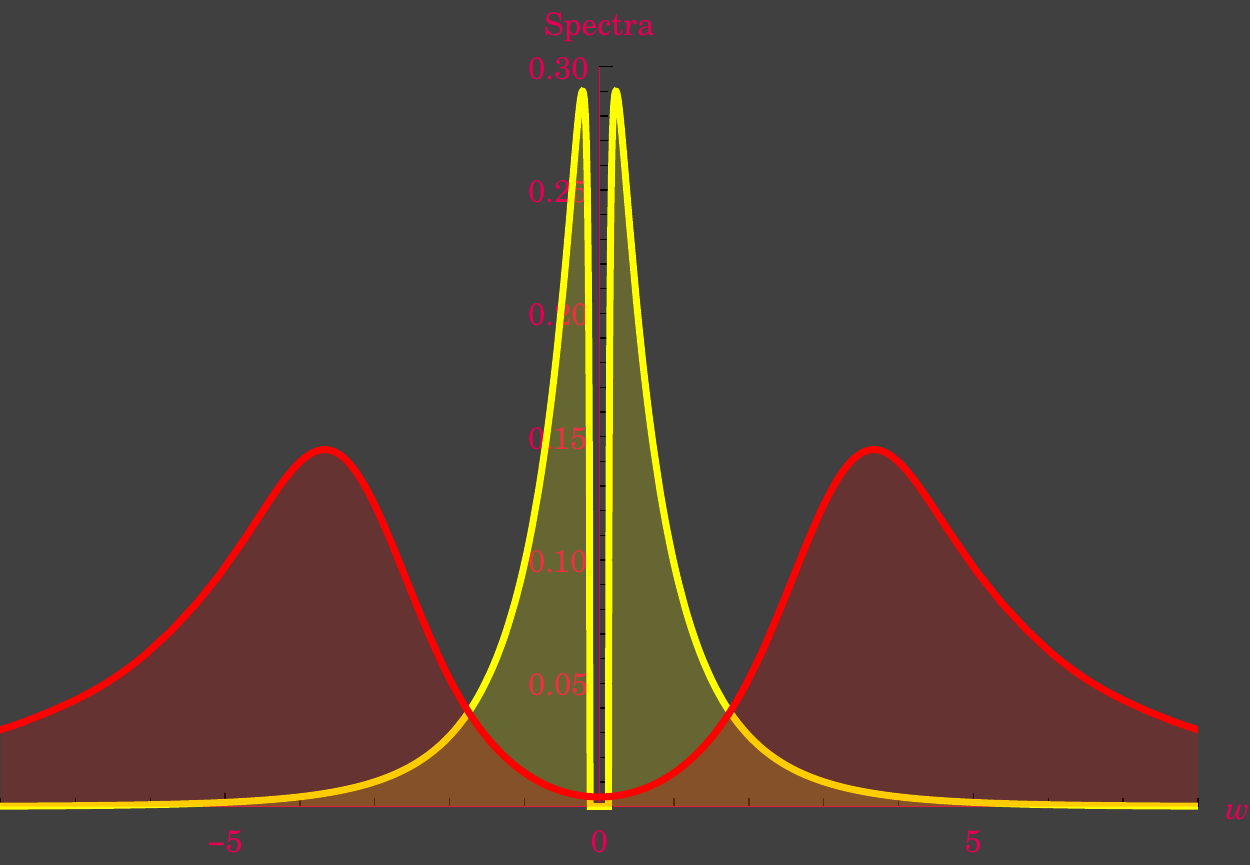}
\caption{Spectral overlap of $\mathcal{G}(\omega)$ (yellow) and $\vert p_{\circ}(t,\omega)\vert^2$ (red) for  $\lambda = -0.01$.}\label{F8}
\end{figure}

\noindent In order to prove that the optimal trajectory presented in Figure (\ref{F6}), offers a physically realizable solution for $p_{\circ}(t)$, we shall calculate the fluence $E$ for this solution. A finite value of $E$ indicates that the solution is physically meaningful. For the solution presented in Figure (\ref{F6}), using the constraint equation (\ref{constraint1}), we numerically estimate $E = 7.03219$, indicating that in this case we indeed have a physically meaningful solution. The spectral overlap of the optimal trajectory in Figure (\ref{F6}), $\vert p_{\circ}(t,\omega)\vert^2$ and the spectrum of the non-adiabatic transitions, $\mathcal{G}(\omega)$ is shown in Figure (\ref{F8})

\noindent The anti-correlation between $\mathcal{G}(\omega)$ and $\vert p_{\circ}(t,\omega)\vert^2$ is clear from Figure (\ref{F8}).

\subsection{Non-adiabaticity of the optimal solution}
Having derived a realistic, optimal solution for the trap velocity, it is imperative to check whether this solution satisfies the adiabatic approximation for the moving Morse potential. A simple, yet physically intuitive check for this can be readily obtained from Figures (\ref{F2}) and (\ref{F8}). We find that the spectrum of non-adiabatic transitions $\mathcal{G}(\omega)$ has non-zero values only for $ \vert\omega \vert > \vert-\omega_0\vert $. Thus, in-order to satisfy the adiabatic approximation, $p_{\circ}(t)$ and hence $p_{\circ}(t,\omega)$ must involve frequencies $\vert \omega \vert \leq \vert -\omega_0 \vert$. However, \,Figure. \ref{F8} shows that the optimal trajectory involves frequencies much larger than $\vert -\omega_0 \vert$, and yet the survival probability is very close to $1$ throughout the transport duration, as illustrated in Figure. \ref{F7}. This is only possible through a formal optimization protocol, that we have followed, whereby, with an explicit knowledge of the leakage spectrum $\mathcal{G}(\omega)$, one can find a trajectory whose spectrum is anti-correlated with the former. In our case $\mathcal{G}(\omega)$ rises to a peak value in $\vert \omega \vert \in [\vert -\omega_0 \,\vert , 1 ]$ and falls off quite rapidly to a negligible amplitude at about $\vert \omega \vert  = 5$ Hz. On the other hand $\vert p_{\circ}(t, \omega) \vert^2$ has peak amplitude at around $\vert \omega \vert = 4$, while having negligible amplitude in $\vert \omega \vert \leq \vert -\omega_0\vert$. Thus we can say that the optimal trap-velocity shown in Figure. \ref{F6}, is not an adiabatic trajectory for the moving Morse potential, for the time-scales under consideration.

\begin{figure}[!h]
\centering
\hspace{0.5cm}\includegraphics[scale= 0.6]{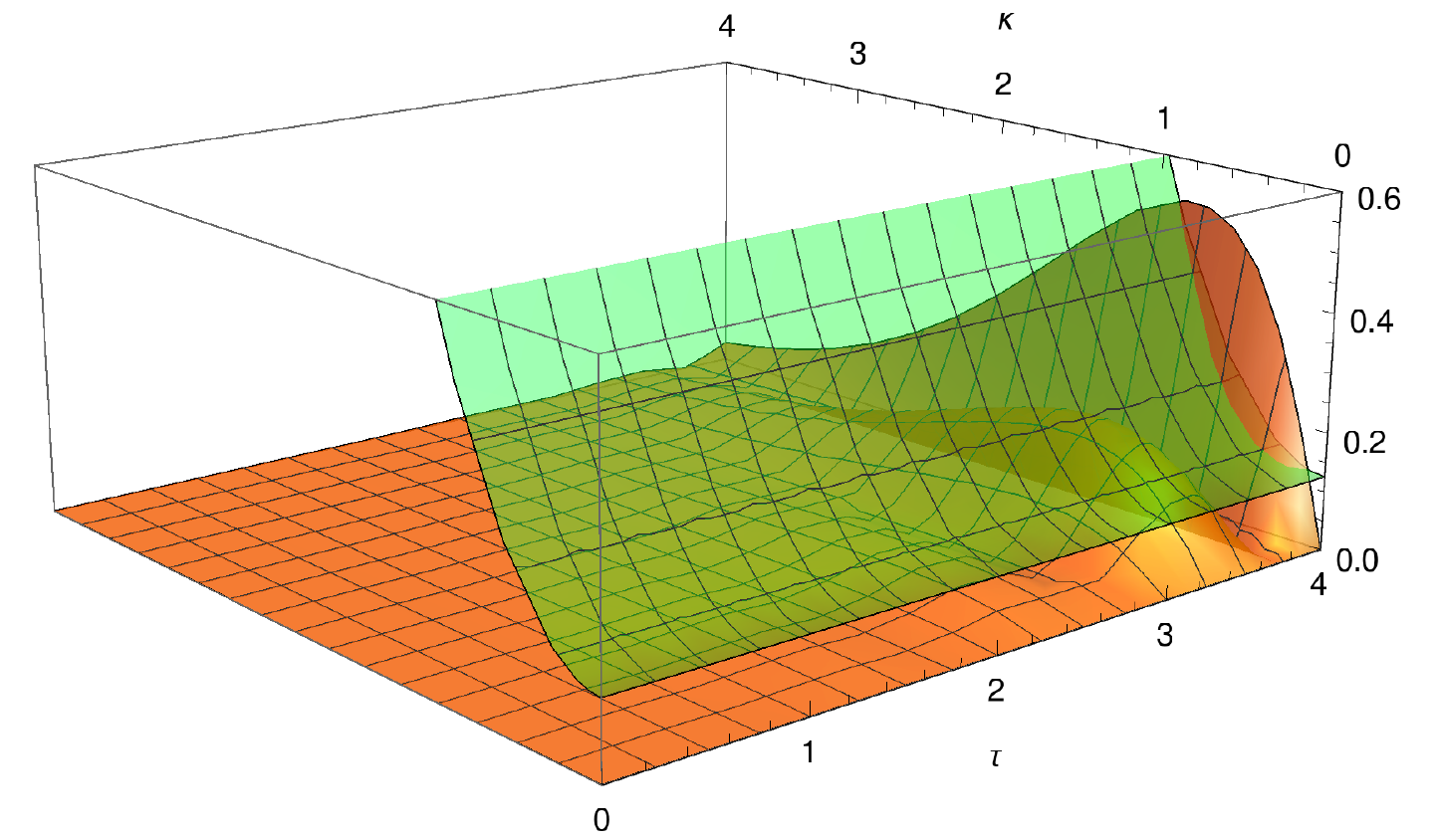}
\caption{Test for adiabaticity of the optimal solution in Fig. \ref{F6}. The orange surface represents $\vert p_{\circ}(\tau)\mu_{0\kappa}\vert$ while the green surface represents $ \vert \omega_{\kappa 0} \vert $. }\label{F9}
\end{figure}

For a more rigorous proof for the non-adiabaticity of the optimal trajectory in Figure. \ref{F6}, we note that in order to satisfy the adiabatic approximation (for the moving Morse potential), we must have \cite{saralidar05}

\begin{eqnarray}
  \underset{0 \leq \tau \leq t}{\mathrm{max}} \;\Bigg\vert\frac{\langle 0(\tau)\vert \dot{H}_I(\tau)\vert \kappa(\tau)\rangle}{\omega_{\kappa 0}}\Bigg\vert \;\ll \; \underset{0 \leq \tau \leq t}{\mathrm{max}}\; \vert \omega_{\kappa 0} \label{AdiabatApprox1}\vert \\
  \implies \underset{0 \leq \tau \leq t}{\mathrm{max}} \; \Big\vert p_{\circ}(\tau)\mu_{0\kappa} \Big\vert \;\ll \; \underset{0 \leq \tau \leq t}{\mathrm{max}}\; \vert \omega_{\kappa 0} \vert , \label{AdiabatApprox2}
\end{eqnarray}
where in the last step we have used equation (\ref{ad3}). In Figure. \ref{F9}, we plot both $\vert p_{\circ}(\tau)\mu_{0\kappa}\vert$ and $ \vert \omega_{\kappa 0} \vert $, as functions of $\tau$ and $\kappa$, to check the validity of (\ref{AdiabatApprox2}) for our optimal solution presented in Figure. \ref{F6}.

\noindent Figure. \ref{F9}, clearly shows that the condition (\ref{AdiabatApprox2}) is not satisfied by the optimal solution, for $\kappa \in \; \sim [0,1]$, thereby establishing its non-adiabaticity.

\subsection{BEC with single Bogoliubov excitation}
We now consider the case where the BEC has a single Bogoliubov excitation at $t = 0$. The generic state of the impurity and the BEC, at any time $t$ is now

\begin{eqnarray}\label{genstate1}
  \vert\psi(t)\rangle & = & \alpha_{\circ}(t)e^{-i\omega_{\circ} t}\vert n(t); 0^B\rangle + \sum_{\kappa}\beta_{\kappa}(t)e^{-\omega_{\kappa} t}\vert \kappa (t) ; 0^B \rangle \nonumber\\
                      &  & + \sum_{k\neq 0} \alpha _{1k}(t) e^{-i(\omega_{\circ} + \Omega_k) t}\vert n(t) ; 1^B_k\rangle \nonumber\\
                      &  & + \sum_{k\neq 0} \sum_{\kappa}\beta _{\kappa 1 k}(t) e^{-i(\omega_{\kappa} + \Omega_k) t}\vert \kappa(t) ; 1^B_k\rangle\nonumber\\
                     &  & + \sum_{k,k'\neq 0} \alpha _{2 k k'}(t) e^{-i(\omega_{\circ} + \Omega_k +\Omega_k') t}\vert n(t) ; 1^B_k, 1^B_{k'}\rangle \nonumber\\ 
                     &  & + \sum_{k,k'\neq 0} \sum_{\kappa}\beta _{\kappa 2 k k'}(t) e^{-i(\omega_{\kappa} + \Omega_k +\Omega_k') t}\vert \kappa(t) ; 1^B_k,1^B_{k'}\rangle,\nonumber\\
\end{eqnarray} 
where as before, we have restricted the analysis to single-phonon exchange sector \cite{Coalson19}. Following the previous analysis, the coefficient vectors for this case are
\begin{equation}\label{Bvector1}
      \bm{B} = 
\begin{bmatrix}
\beta_{\kappa}  \\
\beta_{\kappa 1 k} \\
\beta_{\kappa 2 k k'} \\
\end{bmatrix}\; {\rm and} \; \bm{A} = 
\begin{bmatrix}
\alpha_{\circ}  \\
\alpha_{1k} \\
\alpha_{2 k k'} \\
\end{bmatrix}
\end{equation}
and the corresponding matrices describing the dynamics are $\bm{M} = \begin{bmatrix}
-\gamma_{\kappa\kappa'} & -id_{\kappa\kappa'}^{k*} & 0  \\
-id_{\kappa\kappa'}^k & - \gamma_{\kappa\kappa'} &  -id_{\kappa\kappa'}^{k*} \\
0 & -id_{\kappa\kappa'}^{k*} & 0 \\
\end{bmatrix}$ and $\bm{F} = \begin{bmatrix}
i\gamma_{n\kappa}^{*} & d_{n\kappa}^{k*} & 0  \\
0 & i\gamma_{n\kappa}^{*} & d_{n\kappa}^{k*}\\
0 & 0 & i\gamma_{n\kappa}^{*}\\
\end{bmatrix}$. \\

\noindent With these new definitions, the form of the dynamical equations remain same as in equation (\ref{dyn4}) and the lowest-order solution is given by (\ref{sol3}). If the initial state of the impurity and BEC is $\vert n(0) ; 1_{k_o}^B\rangle$, then the survival probability in the ground state, in the leading order, is given by

\begin{eqnarray}\label{Sprob-11}
  P_n(t) & = & \vert \bm{A}(t) \vert^2 \nonumber\\
         & = &  1 - \int\limits_0^t\, dt_1\int\limits_0^{t_1}\,dt_2\, 2 {\rm Re}\,[G_n(t_1,t_2)],
\end{eqnarray}
where $G_n(t_1,t_2) = d_{n\kappa}^{k}(t_1)d_{n\kappa}^{k*}(t_2) + \gamma_{n\kappa}(t_1)\gamma_{n\kappa}^{*}(t_2)$. As noted before, since $d_{n\kappa}^{k}(t)$ does not depend on $x_{\circ}(t)$, the first term of $G_n(t_1,t_2)$, which represents purely phonon-mediated leakage processes, is not controllable by changing the velocity of the trap. The second term in $G_n(t_1,t_2)$ is same as $\Gamma_n(t_1, t_2)$, which we have analyzed in the previous sections.

An interesting scenario arises when the BEC is initially in a superposition state of the form $\frac{1}{\sqrt{2}}[\vert 0^B\rangle + \vert 1_{k_o}^B\rangle ]$ and the combined initial state is $\frac{1}{\sqrt{2}}[\vert n(0) ;  0^B\rangle + \vert n(0) ; 1_{k_o}^B\rangle ] $. The survival probability in this case becomes

\begin{eqnarray}\label{Sprob-12}
  P_n(t) & = & \frac{1}{2}\Big[ 1 - \int\limits_0^t\, dt_1\int\limits_0^{t_1}\,dt_2\, 2 {\rm Re}\,[\Gamma_n (t_1,t_2) - i \lbrace C_n^{k_o}(t_1, t_2)\rbrace ^{*}]\Big]\nonumber\\
         &  & + \frac{1}{2}\Big[ 1 - \int\limits_0^t\, dt_1\int\limits_0^{t_1}\,dt_2\, 2 {\rm Re}\,[\,i \, C_n^{k_o}(t_1, t_2) + G_n(t_1,t_2)]\Big].\nonumber\\
\end{eqnarray}
Equation (\ref{Sprob-12}) reveals a remarkable, unexpected effect: the survival probability  may additionally depend on the cross-term of non-adiabatic and phonon-mediated processes, through $C_n^{k_o}(t_1, t_2)$, which is controllable by changing the trap velocity. Thus, the optimization in this case should also take into consideration, the cross-spectrum between non-adiabatic and phonon-mediated couplings.

\section{Conclusions}

In this paper we have ventured into an unfathomed domain: the  motion of quantum wavepackets with nonuniform velocity  through dissipative media and the principles of their protection against the adverse effects of this motion and the medium based on a  quantum analysis, from first principles. Remarkable, counter-intuitive quantum friction effects have been revealed by the analytical solutions to this formidable problem in the limit of low temperature, short time and small velocities. The most surprising effect is that dissipation can help preserve the wavepacket intact. This comes about since  the survival probability of the initial state,  as shown, depends not only on the spectra of the non-adiabatic coupling and the phonon bath, but also  on  the cross-spectrum of the non-adiabatic and phonon-mediated transitions. The analytically obtainable optimal trajectory that simultaneously suppresses both transitions  has been shown to be physically sound and feasible. 
 
To conclude, this paper opens a new pathway into a hitherto unexplored field of phenomena combining  quantum many-body effects and the principles of dynamical control that may be highly instrumental in improving our ability to keep quantum friction at bay even under challenging conditions, to the benefit of diverse quantum technological applications.

\section{Acknowledgements}

\noindent GK and IM acknowledge the support of DFG FOR 7024. GK acknowledges the support of PATHOS (FET Open), Pace-In (QUANTERA), ISF and NSF-BSF. AC thanks Biswarup Ash for insightful discussions and suggestions.




\section{Appendix}

\subsection{Eigenfunctions of the Morse Potential}

\begin{equation}\label{diseig}
    \phi_n[z(t)] = \mathcal{N}_n\;[z(t)]^{N-n}\;e^{-z(t)/2}\;M[-n, 2N - 2n + 1, z(t)]
\end{equation}
and
\begin{equation}\label{coneig}
    \phi[\kappa, z(t)] = \mathcal{N}(\kappa)\;[z(t)]^{-i\kappa}\;e^{-z(t)/2}\;U[-N -i\kappa, 1 - 2i\kappa, z(t)],
\end{equation}
where 
\begin{equation}\label{zeig}
    z(t) = (2 N + 1)\,e^{-a[x - x_{\circ}(t)]},
\end{equation}
$M(a,b,z)$ and $U(a,b,z)$ are Kummer functions of first and second kind and

\begin{equation}\label{normn}
      \mathcal{N}_n = \Big[ \frac{(2N - 2n)\Gamma(2N - n + 1)}{n! \Gamma(2N - n + 1)^2}\Big]^{\frac{1}{2}}
\end{equation}
while $\mathcal{N}(\kappa)$ is determined using $\langle \phi[\kappa, z(t)] \vert \phi[\kappa', z(t)]\rangle = \delta(\kappa - \kappa')$, as \cite{lima06, deffner15}
\begin{equation}\label{normk}
      \mathcal{N}(\kappa) = \frac{\vert \Gamma(-N - i\kappa)\vert}{\pi}\sqrt{\kappa \sinh(2\pi\kappa)}.
\end{equation}
It is important to note that both $\phi_n[z(t)]$ and $\phi[\kappa, z(t)]$ are real-valued functions of its arguments
\cite{lima06, deffner15}.

\subsection{Spectrum of Non-adiabatic coupling}
In our problem, we have assumed a single bound-state, so $n = 0$. So,
\begin{equation}\label{w}
     \omega_{n = 0} = \omega_0 = -\frac{a^2}{2 m}(N - 0)^2 = -\frac{N^2}{2 m^*} \hspace{0.25cm}; \hspace{0.25cm} m^* = \frac{m}{a^2}
\end{equation}
and 
\begin{eqnarray}\label{N}
    &(N + \frac{1}{2})^2& = \frac{2 m D}{a^2}\nonumber\\
    &\Rightarrow& N^2 = \frac{2 m D}{a^2} - \frac{\sqrt{2 m D}}{a} + \frac{1}{4}
\end{eqnarray}
\noindent Substituting (\ref{N}) in (\ref{w}) we have,
\begin{equation}
     \omega_{n = 0} = \omega_0 = -\Big[ D - \sqrt{\frac{D}{2 m^*}} + \frac{1}{8 m^*}\Big].
\end{equation}
In order to evaluate $\rm{Re}[\Phi(t)]$ and subsequently $\mathcal{G}(\omega)$ we need an to calculate $\vert \widetilde{\mu}_{n=0,\kappa} \vert^2$, which we shall denote as $\vert \widetilde{\mu}_{0\kappa} \vert^2$. Using the definition of $\widetilde{\mu}_{n\kappa}$ we have

\begin{eqnarray}\label{nonadmu1}
     \widetilde{\mu}_{0\kappa} & = & 2aD\int\limits_{0}^{\infty} \, \frac{dz}{a\,z}\;\phi _0(z)\Big[\Bigg\lbrace \frac{z}{(2 N + 1)}\Bigg\rbrace^2 - \frac{z}{(2 N + 1)}\Big]\phi_{\kappa}(z)\nonumber\\
\end{eqnarray}
where, following equation (\ref{zeig}), we have used the variable substitution $z = (2 N + 1)\,e^{-a q}$. Simplifying, we have

\begin{eqnarray}\label{nonadmu2}
      \widetilde{\mu}_{0\kappa} & = & \frac{2D}{(2 N + 1)^2}\int\limits_0^{\infty} dz \; \phi _0(z)\; z \;\phi_{\kappa}(z) \nonumber\\
             &  &  - \frac{2D}{(2 N + 1)}\int\limits_0^{\infty} dz \; \phi _0(z)\; \phi_{\kappa}(z)\nonumber\\
      & := & \frac{2D}{(2 N + 1)^2}\,I_1 - \frac{2D}{(2 N + 1)}\,I_2
\end{eqnarray}
where in the last step we have defined the first integral on the r.h.s. as $I_1$ and the second integral as $I_2$. Now, using equations (\ref{diseig}), (\ref{coneig}) and (\ref{zeig}) and expressing the Kummer functions of second-kind, $U(a,b,z)$ in terms of the Whittaker's function $W_{\lambda\mu}(z)$ \cite{abramowitz72} we can directly evaluate $I_1$ and $I_2$ using the integrals in \cite{dixit15}, as

\begin{eqnarray}\label{I1-4}
       I_1 & = & \mathcal{N}_0\mathcal{N}(\kappa)\;\Gamma(N + 2 + i\kappa)\Gamma(N + 2 - i\kappa).
\end{eqnarray}
and
\begin{eqnarray}\label{I2-2}
     I_2 & = & \mathcal{N}_0\mathcal{N}(\kappa)\;\Gamma(N + 1 + i\kappa)\Gamma(N + 1 - i\kappa).
\end{eqnarray}

\noindent Substituting (\ref{I1-4}) and (\ref{I2-2}) on the r.h.s. of (\ref{nonadmu2}) we have,

\begin{eqnarray}\label{nonadmufinal}
   \widetilde{\mu}_{0\kappa} & = & \frac{2D\mathcal{N}_0\mathcal{N}(\kappa)}{(2 N + 1)^2}\;\Big[\;\Gamma(N + 2 + i\kappa)\;\Gamma(N + 2 - i\kappa) \nonumber\\
                             &   & - \;(2 N + 1)\;\Gamma(N + 1 + i\kappa)\;\Gamma(N + 1 - i\kappa)\;\Big].
\end{eqnarray}

\subsection{Derivation of the Euler-Lagrange Equation}

The constraint functional is given by

\begin{equation}\label{constraint2}
      J_2[\dot{p}_{\circ}] = \int\limits_0^t \, d\tau\, [\dot{p}_{\circ}(\tau)]^2 - E.
\end{equation}
Introducing a Lagrange multiplier $\lambda$, the total functional to be optimized becomes 

\begin{eqnarray}\label{J}
     J[p_{\circ},\dot{p}_{\circ},t] & = & J_1[p_{\circ}] + \lambda J_2[\dot{p}_{\circ}]\nonumber\\
         & = & \int\limits_0^t\, dt_1\int\limits_0^{t}\,dt_2\,p_{\circ}(t_1)p_{\circ}(t_2)\;\Big\lbrace 2\,\text{Re}[\Phi(t_1 - t_2)]\Big\rbrace \nonumber\\
         &  &  \hspace{2cm}  + \;\lambda\Big[\int\limits_0^t \, d\tau\, [\dot{p}_{\circ}(\tau)]^2 - E \Big]
\end{eqnarray}
Since the double integral in the first term on the r.h.s. of (\ref{J}) extends over a full square, we can use the results of Non-local Variational Mechanics, derived in reference \cite{edelen69}, to obtain the first variation of $J$ as

\begin{eqnarray}\label{Jvar}
   \delta J & = & 2\int\limits_0^t\,dt_1\, \delta p_{\circ}(t_1)\,\int\limits_0^{t}\,dt_2\,p_{\circ}(t_2)\;\Big\lbrace 2\,\text{Re}[\Phi(t_1 - t_2)]\Big\rbrace \nonumber\\
            &   & - 2\lambda \int\limits_0^t \, dt_1 \, \delta p_{\circ}(t_1)\, \ddot{p}_{\circ}(t_1).
\end{eqnarray}
The stationarity condition $\delta J = 0$ then results in the Euler-Lagrange Equation.

\subsection{The Lagrange multiplier}

From equation (\ref{constraint1}) we have 

\begin{equation}\label{l1}
      \int\limits_0^t \, d\tau_1\, \Big[\, \int\limits_0^{\tau_1} \, d\tau_2 \; \ddot{p}_{\circ}(\tau_2)\,\Big]^2 = E.
\end{equation}
Then using equation (\ref{ELeq}) we have

\begin{equation}\label{l2}
     \frac{1}{\lambda^2} \int\limits_0^t \, d\tau_1\, \Big[\, \int\limits_0^{\tau_1} \, d\tau_2 \; \int\limits_0^{\tau_2}\, d\tau_3 \; p_{\circ}(\tau_3)\;2\,\text{Re}\lbrace\Phi(\tau_2 - \tau_3)\rbrace\,\Big]^2 = E.
\end{equation}
Solving for $\lambda$ we then have

\begin{equation}\label{l3}
    \lambda = \pm \frac{1}{\sqrt{E}}\sqrt{\int\limits_0^td\tau_1\Big[\int\limits_0^{\tau_1}d\tau_2\int\limits_0^{\tau_2}d\tau_3\, p_{\circ}(\tau_3)\;2\text{Re}\lbrace\Phi(\tau_2 - \tau_3)\rbrace\,\Big]^2}.
\end{equation}
Now, since $p_{\circ}(t) \in \mathbb{R}$, equation (\ref{constraint1}) implies that $E > 0$, for non-trivial $p_{\circ}(t)$. Equation (\ref{l3}) then implies that $\lambda \in \mathbb{R}$.

\end{document}